\documentclass{aa}
\usepackage{psfig}

\usepackage{times}
\usepackage{mathptm}
\fontfamily{ptmcm}\selectfont

\def\subr #1{_{{\rm #1}}}
\def\etal{{et al.} }

\def\gradip{\hbox{\rlap{\hbox{.}}\raise 5.truept \hbox{{\small $\circ$}}}}
\def\cf{{\em cf.\/}}

\begin{document}

\thesaurus{08.05.3; 
           08.08.1; 
           08.12.3; 
           08.16.3; 
           10.07.3 NGC~6254, NGC~6656, NGC~6809}

\title{HST Luminosity Functions of the Globular Clusters 
       M10, M22, and M55. A comparison with other clusters.
\thanks{Based on HST observations retrieved from the ESO ST-ECF
        Archive, and on observations made at the European Southern
        Observatory, La Silla, Chile, and at the JKT telescope at 
        La Palma, Islas Canarias.}}

\author{G. Piotto \inst{1} \and M. Zoccali \inst{1} }

\offprints{G. Piotto, e-mail: piotto@pd.astro.it}

\institute{Dipartimento di Astronomia -- Universit\`a di
           Padova, Padova, ITALY }

\date{Received ; accepted }
\titlerunning{Globular Cluster Luminosity and Mass Functions}
\maketitle
\markboth{Piotto and Zoccali}{HST LFs of M10, M22, and M55}

\begin{abstract}
From a combination of deep Hubble Space Telescope $V$ and $I$ images
with groundbased images in the same bands, we have obtained
color--magnitude diagrams of M10, M22, and M55, extending from just
above the hydrogen burning limit to the tip of the red giant branch,
down to the white dwarf cooling sequence. We have used the
color-magnitude arrays to extract main sequence luminosity functions
(LFs) from the turnoff to $m \sim 0.13m_\odot$. The LFs of M10 is
significantly steeper than that for the other two clusters. The
difference cannot be due to a difference in metallicity.  A comparison
with the LFs from Piotto, Cool, and King (1997), shows a large spread
in the LF slopes. This spread is also present in the local mass
functions (MFs) obtained from the observed LFs using different
theoretical mass--luminosity relations. The dispersion in the MF
slopes remains also after removing the mass segregation effects by
using multimass King-Michie models.  The globular cluster MF slopes are
also flatter than the MF slope of the field stars and
of the Galactic clusters in the same mass interval. We interpret the
MF slope dispersion and the MF flatness as an evidence of dynamical
evolution which makes the present day globular cluster stellar MFs
different from the intial MFs.  The slopes of the present day MFs
exclude that the low mass star can be dynamically relevant for the
Galactic globular clusters.
 
\end{abstract}


\section{Introduction} \label{intro}

The {\it Hubble Space Telescope (HST)} allows the derivation of color
magnitude diagrams (CMDs) of Galactic globular clusters (GCs) which
extend to almost the faintest visible stars, just above the
hydrogen-burning limit for the nearest clusters (King \etal\ 1998). 
These CMDs can be used to extract luminosity functions
(LFs), and from them mass functions (MFs) which extend over almost the
entire mass range of the luminous GC stars, from the turnoff (TO)
down to the bottom of the main sequence ($0.1\leq m/m_\odot\leq0.8$).

The GC MFs are important as they can provide important observational
inputs in a variety of astrophysical problems, like the realistic
dynamical modeling of individual clusters (King, Sosin, and Cool 1995,
Sosin 1997), and the role of dynamical evolution in modifying the GC
stellar content (Piotto, Cool, and King 1997, PCK). In principle, GC MFs
also give information about the amount of mass contained in very-low-mass 
stars and
brown dwarf stars in globulars, and, by extension, in the Galactic
halo. The observed MFs are related to the GC initial MFs (IMFs): a basic input
parameter for any GC and galaxy formation model.

Progress on these issues requires accurate photometry of main sequence
(MS) stars. In many cases, the HST data alone cannot provide all the
information. The brightest MS stars in the closest GCs are always
badly saturated in deep WFPC2/HST frames, and often there are 
no short exposures in the same field, as we tend to use HST to measure the
faintest objects. This lack of information might become a dangerous
drawback when comparing the stellar LF of different clusters, as
discussed in Cool, Piotto, and King (1996, CPK). In fact, the shape of the
LF for stars fainter than $M_V\sim8$ is dominated by the slope of the
mass-luminosity relation (MLR) more than by the shape of the MF.
Ground-based data might become of great importance in this respect.

Here we present the first deep HST CMDs and LFs for M10 (NGC~6254) and
M55 (NGC~6809), and an independent determination of the CMD and the LF
of M22 (NGC~6656).  A CMD and a LF of M22 have already been presented
by De Marchi and Paresce (1997), based on the same images.  For all
three clusters, the CMDs and LFs have been extended to the TO and
above, by means of ground-based data.  We compare the results for M22
with the LF of De Marchi and Paresce (1997), and then compare the LFs
for the three clusters with each other.  A comparison with the
presently available HST LFs and MFs for intermediate and metal-poor
clusters is also presented. This paper is a continuation and an
extension of the paper by PCK.

A description of the observations and of the data analysis is presented
in Section~\ref{dataset}. The CMDs and LFs are presented in Sections \ref{cmd}
and \ref{lf}. The MF is derived in Section~\ref{mf}. A comparison of
the presently available LFs extended to the TO is shown in Section ~\ref{comp},
and a discussion follows in Section~\ref{disc}.

\begin{table}
\caption[]{Data Set}
\label{log}
\begin{tabular}{ccccr}
\hline
Object & Telescope & Obs. date & Filter & Exp. time (s) \\
\hline
 M22   & HST   & 30-9-1995  & F606W & 1100          \\
       &  ''   &  ''        & F606W & 1200$\times$3 \\
       &  ''   &  ''        & F814W & 1100          \\
       &  ''   &  ''        & F814W & 1200$\times$3 \\
       & DUTCH & 15-4-1997  & V     & 45            \\
       &   ''  &  ''        & V     & 1500          \\
       &   ''  &  ''        & I     & 45            \\
       &   ''  &  ''        & I     & 1500          \\
&&&& \\
M55    & HST   & 4-11-1995  & F606W & 1100          \\
       &  ''   &   ''       & F606W & 1200$\times$5 \\
       &  ''   &   ''       & F814W & 1100          \\
       &  ''   &   ''       & F814W & 1200$\times$5 \\
       & DUTCH & 15-4-1997  & V     & 45            \\
       &  ''   &   ''       & V     & 1500          \\
       &  ''   &   ''       & I     & 45            \\
       &  ''   &   ''       & I     & 1500          \\
&&&& \\
 M10   & HST   & 10-10-1995 & F606W & 1100          \\
       & ''    &   ''       & F606W & 1200$\times$9 \\
       & ''    &   ''       & F814W & 1100          \\
       & ''    &   ''       & F814W & 1200$\times$9 \\
       & JKT   & 30-5-1997  & V     & 15            \\
       & ''    &   ''       & V     & 45            \\
       & ''    &   ''       & V     & 1500          \\
       & ''    &   ''       & I     & 15            \\
       & ''    &   ''       & I     & 45            \\
       & ''    &   ''       & I     & 1500          \\
\hline
\end{tabular}
\end{table}

\section{Observations and Analysis} \label{dataset}
The data presented in this paper consist of a set of WFPC2 images and
a set of ground-based frames covering approximately the same field for
the three GCs M10, M22, and M55.

The WFPC2 data have been taken from the ST-ECF HST archive and were
obtained with the F606W and F814W filters during Cycle 5.  
The observation dates and the exposure times are given in
Table~\ref{log}. 
The fields are located at about 3.0 arcmin (for M10), 4.5 arcmin (for M22),
and 3.0 arcmin (for M55) from the cluster centers.
The HST archive data did not include short exposures. This fact made
impossible the photometry of any stars brighter than V$\sim19.0$,
because of the CCD saturation problems.
This means that the CMDs and the LFs from the HST photometry are truncated
at about two magnitudes below the TO. 

In order to extend both the CMDs and the LFs to the TO and above, we
collected a set of short-exposure images with three 1-m size
telescopes. For M22 we used the 0.9m Dutch telescope at ESO (La
Silla), the 1.54m Danish telescope at ESO for M55, and the 1.0m
Jacobus Kapteyn Telescope (JKT) at La Palma (Islas Canarias) for
M10 ({\it cf.}  Table~\ref{log} for the observation dates and the
exposure times). The seeing conditions were exceptionally good at the
JKT (0.7 arcsec FWHM), while the average seeing conditions for the ESO
images were
1.2 arcsec FWHM. The ground-based images for M10 and M22 were
collected in photometric conditions, while the M55 data come
from a non-photometric night.

\subsection{HST data reduction}\label{hstdata}
All HST observations were pre-processed through the standard HST
pipeline with the most up-to-date reference files. Following
Silbermann \etal\ (1996), we have masked out the vignetted pixels,
saturated and bad pixels and columns using a vignetting mask created
by P.B. Stetson together with the appropriate data quality file for
each frame. We have also multiplied each frame by a pixel area map
(also provided by P.B. Stetson) in order to correct for the geometric
distortion (Silbermann \etal\ 1996).

The photometric reduction was carried out using the\\
DAOPHOT~II/ALLFRAME package (Stetson 1987, 1994).  A preliminary
photometry was carried out in order to construct an approximate list
of stars for each single frame. This list was used to accurately match
the different frames.  With the correct coordinate transformations
among the frames, we obtained a single image, combining all the
frames, regardless of the filter. In this way we could eliminate all
the cosmic rays and obtain the highest signal/noise image for star
finding.  We ran the DAOPHOT/FIND routine on the stacked image and
performed PSF-fitting photometry in order to obtain the deepest list
of stellar objects free from spurious detections. Finally, this list
was given as input to ALLFRAME, for the simultaneous PSF-fitting
photometry of all the individual frames. The PSFs we used 
were the WFPC2 model PSFs extracted by P.B. Stetson (1995) from
a large set of uncrowded and unsaturated images.

We transformed the F606W and F814W instrumental magnitudes into the
standard $V$ and $I$ systems using Eq. (8) of Holtzman \etal\ (1995) and
the coefficients in their Table~7.

Note that the CMDs of M10, M22, and M55 presented in the following
come from the combination of the photometry in the three WF chips.
The LFs have been obtained from chip WF2 for M10, from WF3 for
M22, and WF4 for M55. These fields contain the largest number of
unsaturated stars and the smallest number of saturated pixels.  In
view of the small error bars of the LFs presented in Section~\ref{lf},
we considered it not worth the large cpu time that would have been
required to run crowding experiments for every chip.

\subsection{Ground-based data reduction}\label{gbdata}
Image pre-processing (bias subtraction and flatfielding) was carried
out using standard IRAF routines. The stellar photometry has been obtained
using DAOPHOT~II/ALLFRAME as described above on all the images (including 
the short exposures) simultaneously. We constructed the model PSF for each
image using typically $\sim$ 120 stars. 

Since some of the ground-based images have been collected on 
non-photometric nights, the calibration of the instrumental magnitudes was
performed by comparison with the stars in the overlapping HST
fields. We first adopted the color term obtained for the same
telescopes during the previous nights in the same run (Rosenberg
\etal\ 1999). 
The zero points have been calculated by comparing all the
non-saturated stars in the WFPC2 chips that were also measured in the
ground-based images. The uncertainties in the $V$ zero points are
0.004, 0.035, and 0.015 magnitudes (the errors refer to the errors on the
mean differences) for M10, M22, and M55, respectively.  The
errors in the $(V-I)$ colors are 0.006, 0.055, and 0.025, respectively.
The zero point uncertainties for M22 are noticeably larger than for the other
two clusters. This fact is due to the smaller overlap in magnitude
between the ground-based and the HST photometries, as can be seen
in Fig.~\ref{cm22}.  The calibration of the M22 and M55 CMDs has been
further checked by comparing these diagrams with other independently
calibrated $V$ vs. $(V-I)$ ground-based CMDs kindly provided by Alfred
Rosenberg 
(Ro\-sen\-berg \etal\ 1999). The two sets of data are consistent within the 
uncertainties given above.

Shorter-exposure HST images (or longer-exposure ground-based frames) are
desirable for a smoother overlap of the two data sets. Note that this
problem does not affect the LF presented in Section~\ref{lf}. Indeed, a
zero point error of a few hundreds of a magnitude is perfectly
acceptable for a LF with magnitude bins of 0.5 mag.

\subsection{Artificial star tests}\label{astest}

Particular attention was devoted to estimating the completeness of our
samples.  The completeness corrections have been determined by
standard artificial-star experiments on both the HST and ground-based
data. For each cluster, we performed ten independent experiments for
the HST images and five for the ground-based ones. In order to
optimize the cpu time, in our experiments we tried to add the largest
possible number of artificial stars in a single test, without
artificially increasing the crowding of the original field.  The
artificial stars have been added in a spatial grid such that the
separation of the centers in each star pair was two PSF radii plus one
pixel. The position of each star is fixed within the grid. However,
the grid was randomly moved on the frame in each different
experiment. We verified that in this way we were not creating
over-crowding by running an experiment with half the number of
artificial stars.  The finding algorithm adopted to identify and
measure the artificial stars was the same used for the photometry of
the original images.  The artificial stars were added on each single V
and I frame.  For each artificial star test, the frame to frame
coordinate transformations (as calculated from the original
photometry) have been used to ensure that the artificial stars were
added exactly in the same position in each frame. We started by adding
stars in one V frame at random magnitudes; the corresponding I
magnitude for each star was obtained using the fiducial line of the
instrumental CMD.  Finally, in each band, we scaled the magnitudes
according to the frame to frame magnitude offset as calculated from
the original photometry.  The frames obtained in this way were stacked
together in order to perform star finding and obtain the most complete
star list. The latter was used to reduce the single frames
simultaneously with ALLFRAME, following all the steps and using the
same parameters as on the original images.

In order to take into account the effect of the migration of stars
toward brighter magnitudes in the LF (Stetson \& Harris 1988), we
corrected for completeness using the matrix method described in
Drukier \etal\ (1988).

In the LFs presented here, we include only points for which the
completeness figures were 50\% or higher, so that none of the counts
have been corrected by more than a factor of 2.

A comparison between the added magnitudes and the measured magnitudes
allows also a realistic estimate of the photometric error
$\sigma\subr{pho}$ (defined as the standard deviation of the differences
between the magnitudes added and those found) as a function of 
magnitude. We use this information in different places in what follows.

\begin{figure}
\psfig{figure=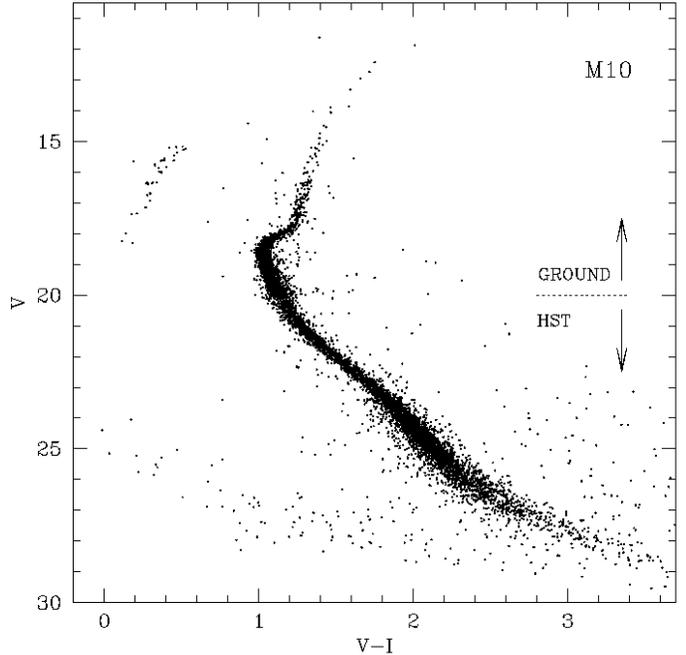,width=9.5cm}
\caption[]{Composite CMD for 6986 stars in M10. 
The ground-based data are from the JKT telescope.}
\label{cm10}
\end{figure}

\section{The color-magnitude diagrams} \label{cmd}

\begin{figure}
\psfig{figure=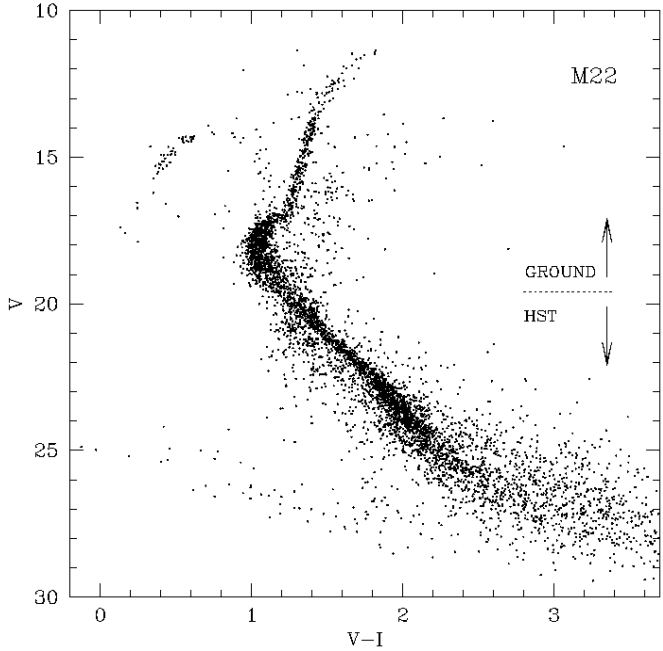,width=9.2cm}
\caption[]{Composite CMD of 5385 stars in M22. 
The ground-based data are from the ESO-Dutch
telescope.  For the HST data only the stars in the WF2 field are shown.}
\label{cm22}
\end{figure}

\begin{figure}
\psfig{figure=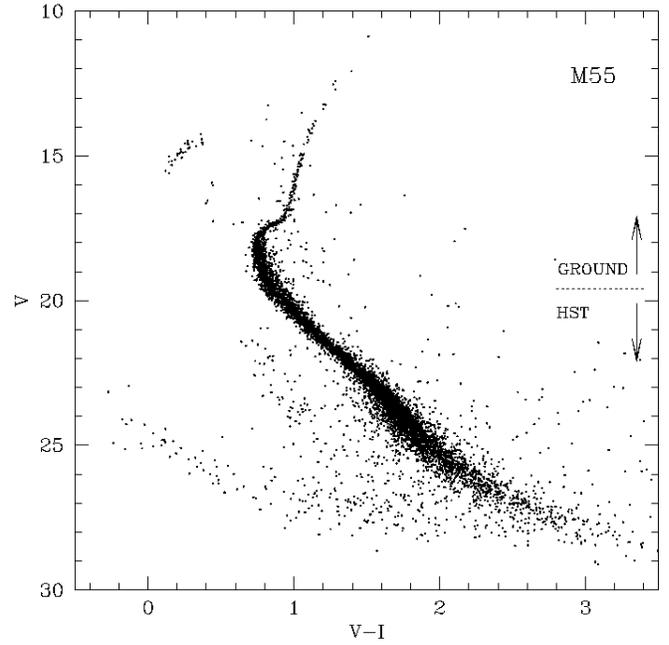,width=9.2cm}
\caption[]{Composite CMD of 8121 stars in M55. 
The ground-based data are from the ESO-Danish telescope.}
\label{cm55}
\end{figure}

\begin{figure}
\psfig{figure=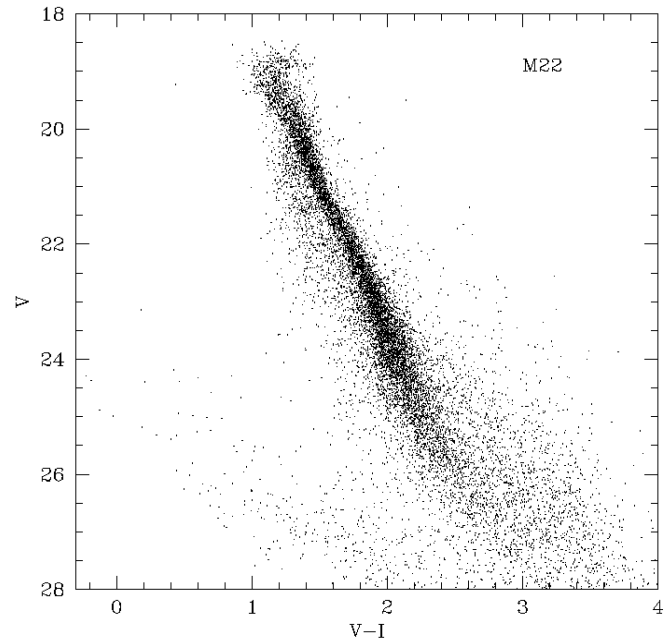,width=9.2cm}
\caption[]{CMD of 13359 stars from the three WF fields of M22. 
The large dispersion of the MS is interpreted in terms of a
differential reddening of $\sim0.05$ magnitudes in $(V-I)$.}
\label{cm22all}
\end{figure}

The CMDs derived from the photometry discussed in the previous Section
are presented in Figs.~\ref{cm10}, \ref{cm22}, and \ref{cm55}.
The upper part of the CMDs comes from the ground-based data, 
while the lower part is from the three WF
cameras of the WFPC2. In the case of M22, the CMD for magnitudes
fainter than V=19.8 comes from the WF2 only: the differential
reddening of this cluster (Peterson and Cudworth 1994) makes the
sequence much broader than expected from the photometric errors.  The
MS of M22 from the three WF cameras is shown in Fig.~\ref{cm22all}.
In Table~\ref{disp} the dispersion $\sigma_{MS}$ (defined as the sigma
of the best fitting gaussian) of the MS (after the removal of the
field star contamination as described in Section~\ref{lf}) is compared
with the expected photometric error $\sigma_{(V-I)}$. The latter have
been estimated from the artificial star tests (\cf\
Section~\ref{astest}), in one magnitude bins, in the interval
$20<V<26$. The resulting average differential reddening in the 3 WF
fields is
$\langle\sigma\subr{red}\rangle=\big\langle\sqrt{\sigma\subr{MS}^2-\sigma_{(V-I)}^2}\big\rangle=0.05$
magnitudes. This value must be considered as an upper limit for the
differential reddening in this region.

\begin{table}
\caption[]{}
\label{disp}
\[
\begin{array}{llll} 
\hline
\noalign{\smallskip}
 V & \sigma_{MS} & \sigma_{(V-I)} & \sigma_{red} \\
\noalign{\smallskip}
\hline
\noalign{\smallskip}
20.5   &   0.068  &  0.036   &   0.058 \\
21.5   &   0.069  &  0.045   &   0.052 \\
22.5   &   0.072  &  0.052   &   0.048 \\
23.5   &   0.081  &  0.071   &   0.039 \\
24.5   &   0.093  &  0.079   &   0.049 \\
25.5   &   0.100  &  0.086   &   0.051 \\
\noalign{\smallskip}
\hline 
\end{array}
\] 
\end{table}

The ground-based and the HST fields are partially overlapping, with
the ground-based images always covering a larger portion of the
cluster.  A detailed discussion of these CMDs will appear
elsewhere. Here it suffice to note that we measured stars from the tip
of the giant brach to a limiting magnitude $V\sim28$. A white dwarf
cooling sequence is clearly seen in all diagrams (but it will be
discussed elsewhere). For the first time, we have a complete picture
of a simple stellar population about 15 Gyr after its birth, from
close to the hydrogen-burning limit to the final stages of its
evolution along the white dwarf sequence. These diagrams can be used
for a fine tuning of the stellar evolution and population synthesis
models (Brocato \etal\ 1996).

Contamination by foreground/background stars is small for M10, as
expected from its galactic latitute ($b=23^\circ$), though a few background
stars (likely from the outskirts of the Galactic bulge) are present.
Despite the fact that M55 has the same latitude as M10, a significantly
larger fraction of field stars is visible in the CMD of
Fig.~\ref{cm55}. Some of these stars are likely bulge members, but the
prominent sequence blueward of the MS of M55 must be associated with the
MS and TO of the stars in the Sagittarius dwarf spheroidal galaxy
(Mateo \etal\ 1996, Fahlman \etal\ 1996).  M22 is the most
contaminated cluster. Both Galactic disk and Galactic bulge stars are
clearly seen in the CMDs of Figs.~\ref{cm22} and \ref{cm22all}.

Deep CMDs also contain information on the low-mass content of the 
clusters.  This information can be extracted from our data only after
we have a reliable transformation from luminosities to
masses. Unfortunately, such a transformation remains uncertain for
low-metallicity, low-mass stars. Almost nothing is known from the
empirical point of view, and different calculations of stellar models
yield different masses, particularly for the lowest-mass stars (King
\etal\ 1998), and different overall trends (slopes) for the
mass-luminosity relations (MLRs).

\begin{figure}
\psfig{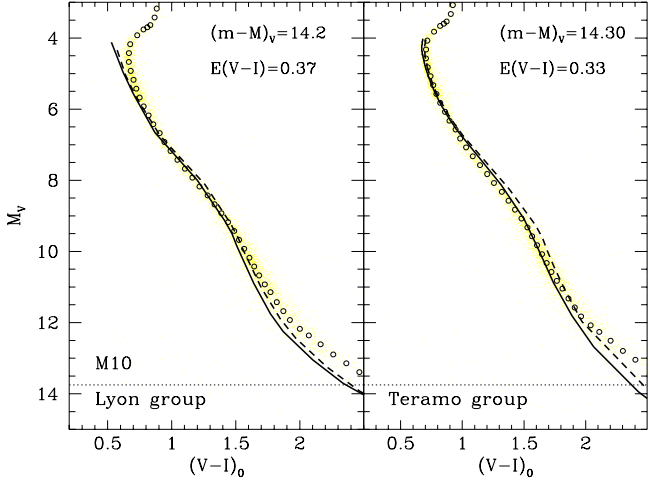}
\caption[]{Comparison between the observed CMD of M10 and the
models by the Lyon group ({\it left panel}), and the Teramo group
({\it right panel}) for [M/H] $=-1.3$ ({\it solid line}) and [M/H] $=-1.5$
({\it dashed line}). The {\it open circles} represent the MS ridgeline.}
\label{m10teo}
\end{figure}

\begin{figure}
\psfig{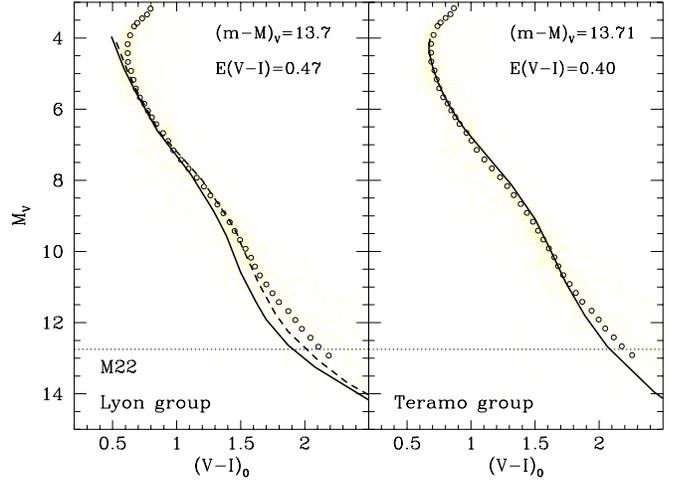}
\caption[]{Comparison between the observed CMD of M22 and the
models by the Lyon group ({\it left panel}), and the Teramo group
({\it right panel}) for [M/H] $=-1.5$ ({\it dashed line}) and [M/H] $=-2.0$
({\it solid line}). The {\it open circles} represent the MS ridgeline.}
\label{m22teo}
\end{figure}

\begin{figure}
\psfig{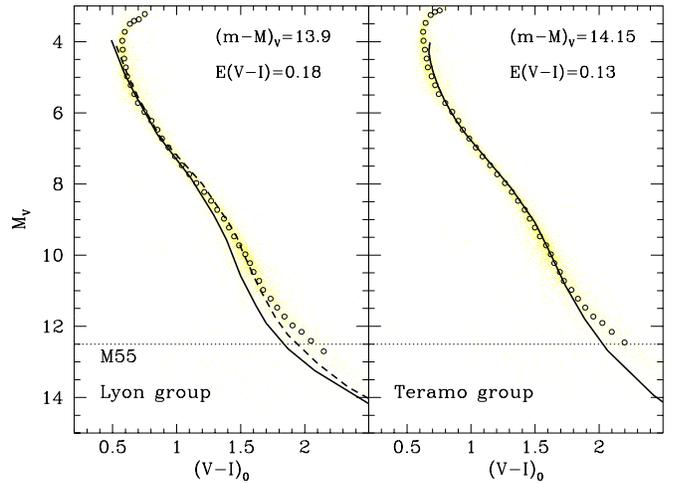}
\caption[]{Comparison between the observed CMD of M55 and the
models by the Lyon group ({\it left panel}), and the Teramo group
({\it right panel}) for [M/H] $=-1.5$ ({\it dashed line}) and [M/H] $=-2.0$
({\it solid line}). The {\it open circles} represent the MS ridgeline.}
\label{m55teo}
\end{figure}

As already found for NGC~6397 (King \etal\ 1998) and the other three
metal-poor clusters studied by PCK (\cf\ their Fig.~3), among the 
existing models we find that those by the group in Lyon (Baraffe
\etal\ 1997) and by the group in Teramo (Cassisi \etal\ 1998, in 
preparation) best reproduce the observed sequences of M10, M22, and
M55.  [Note that Cassisi et al.'s (1998) models below $0.5m_\odot$
($M_V\sim8.1$) are the same models as in Alexander \etal\ (1997).]
The level of agreement between the models and the observed data can be
fully appreciated in Figs. \ref{m10teo}, \ref{m22teo}, and
\ref{m55teo}. In these figures the open circles represent the MS
ridgeline, obtained by using a mode-finding algorithm and a
kappa-sigma iteration in order to minimize the field star
contamination. The dotted line represents the $V$ magnitude limit of
the LFs presented in the following Section~\ref{lf}; the data below
this magnitude limit are not used in the present paper.  The dashed
line shows the isochrone corresponding to the metallicity which best
matches the Zinn and West (1984) [Fe/H] (iron) content, scaled to the
appropriate metallicity [M/H] assuming [O/Fe]=0.35 (Ryan and Norris
1991), and using the relation by Salaris, Chieffi, and Straniero
(1993).  According to Table~\ref{distance}, we used the models for
[M/H] $=-1.5$ for M22 and M55, and the models for [M/H] $=-1.3$ for
M10.  For comparison reasons, in Figs.\ref{m10teo}, \ref{m22teo}, and
\ref{m55teo} we show also the isochrones which best match the
Zinn and West (1984) metallicity assuming a solar ratio for the
alpha elements (solid line).  The distance modulus and reddening have
been left as free parameters. The resulting values of $(m-M)_V$ are in
very good agreement with the values in the literature (\cf\ Djorgovski
1993). This is also true for the $E(V-I)$ resulting from the fit of the
Lyon group models. The models from the Teramo group result
systematically redder by about 0.06 magnitudes in $(V-I)$ than the
isochrones from Baraffe \etal\ (1997), and the resulting reddening is
marginally consistent with the reddening in the literature.  
In the LF comparison discussed in Section~\ref{lf},
we have adopted the distance moduli and reddenings 
used in the fit of the Baraffe \etal\ models and listed in 
Table~\ref{distance}. 

We want to briefly comment on the comparisons in
Figs.\ref{m10teo}, \ref{m22teo}, and \ref{m55teo}, leaving a more complete
discussion to a future paper specific to the CMDs. There is an overall
agreement between the models and the observed sequences.  With the adopted
distance moduli, both sets of
models reproduce the characteristic bends of the MS, and at the
correct magnitudes. The MSs of M22 and M55 seem to be better
reproduced by the models, while the discrepancies seems to be more
significant for M10.  The deviations close to the TO might be due to
the age of the adopted isochrones (the only ones available to us),
which is 10 Gyr for the Lyon models and 14 Gyr for the Teramo
ones. The isochrones seem to deviate more and more in color in
the lowest part of the CMD. We can exclude that this is due to any
internal errors in our photometry. The artificial-star experiments
show that the average deviation in color due to photometric errors
is less than 0.03 magnitudes at the faintest limit of the photometry.
The residual differences might arise both from errors in the
calibration from the HST to the standard $(V,I)$ system and to errors in
the transformation from the theoretical to observational plane, very
uncertain for these cool stars (Alexander \etal\ 1997).


\begin{table}
\caption[]{Adopted Parameters}
\label{distance}
\begin{tabular}{llllll}
\hline
Object & model & $(m-M)_V$ & E(B-V) & [M/H]  \\
\hline
&&&&\\
M10    & Baraffe et al. & 14.20  &  0.29 &  -1.3  \\
       & Cassisi et al. & 14.25  &  0.23 &  -1.3  \\
&&&&\\
M22    & Baraffe et al. & 13.70  &  0.37 &  -1.5  \\
       & Cassisi et al. & 13.71  &  0.31 &  -1.54 \\
&&&&\\
M55    & Baraffe et al. & 13.90  &  0.14 &  -1.5  \\
       & Cassisi et al. & 14.15  &  0.10 &  -1.54 \\
&&&&\\
	
\hline
\end{tabular}
\end{table}


\section{The luminosity functions} \label{lf}

The principal objective of the present work is to measure 
main-sequence luminosity functions. To this end, we followed the same
procedure outlined in CPK and PCK.  We
started by locating the ridge line of the MS using a kappa-sigma
clipping algorithm both on the ground-based and on the HST data. The mode of
the $(V-I)$ distribution was calculated in bins 0.25 mag wide in $V$.
The next step was to subtract the MS ridge-line color from the measured
color of each star, in order to produce a CMD with a straightened MS.
Two lines, to the left and right of the straightened sequence, were
then used to define an envelope around the MS. A range of $\pm3
\sigma$ (where $\sigma$ is the standard deviation of the photometric
errors obtained from the artificial-star experiments) was used in
order to encompass nearly all the MS stars. Stars within this
envelope were binned in 0.5 mag intervals in order to produce a
preliminary LF.  These LFs must be corrected for field-star
contamination. The field-star density was estimated by taking the
average of the star counts in strips of width $6 \sigma$ outside each
side of the MS envelope. In M10 and M55 the correction for field stars
was always less than 10\%.  

The case of M22 is more complicated. The field-star contamination is
below 10\% only in the magnitude interval $21.5<V<26$. At $V=26.25$
the contamination is already 15\%.  At fainter magnitudes, it becomes
rather uncertain because of the spread of the MS due to the
photometric errors. We think that below $V=26.5$ ($I=23.8$) it is not
possible to estimate the field star contamination reliably with the
present data for M22. In the magnitude interval $19.5<V<21.0$ the red
giant branch of the bulge crosses the MS of M22, creating additional
problems for the estimate of the field-star contamination. In this
magnitude interval we estimated the amount of contamination using the
counts obtained running the code by Ratnatunga and Bahcall
(1985), normalized to the number of field stars we found in
the box defined by $21.5<V<20.0$ and $1.0<(V-I)<1.5$ in the CMD of
Fig.~\ref{cm22}.
For this cluster, a more reliable LF can be obtained only with
second-epoch HST observations of the same field analyzed in this paper,
as was done by King \etal\ (1998) for NGC 6397. 

\begin{figure}
\psfig{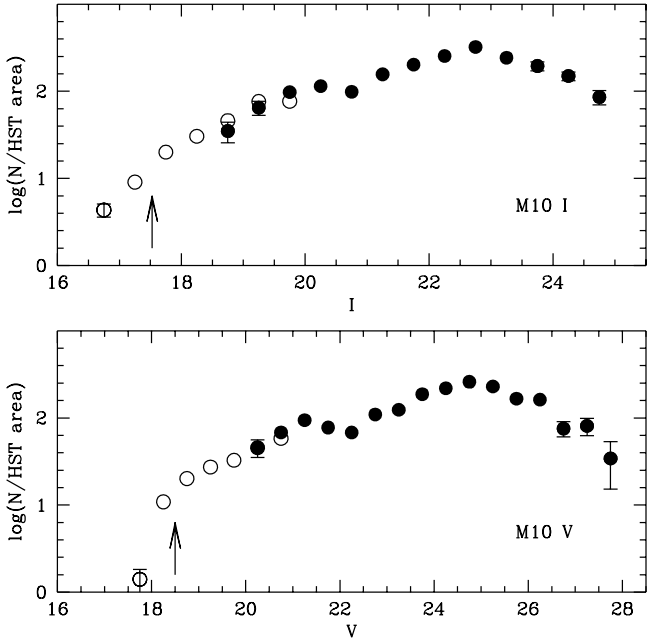}
\caption[]{The $V$ ({\it lower panel}) and $I$ ({\it upper panel}) 
ground-based ({\it open circles}) and HST ({\it filled circles}) LFs 
of M10 from the CMD of Fig.~\ref{cm10}. The ground-based LFs have been 
scaled to the HST area. The arrows indicate the TO position. The error 
bars are plotted only when they exceed the symbol size.}
\label{m10_gb_hst}
\end{figure}

\begin{figure}
\psfig{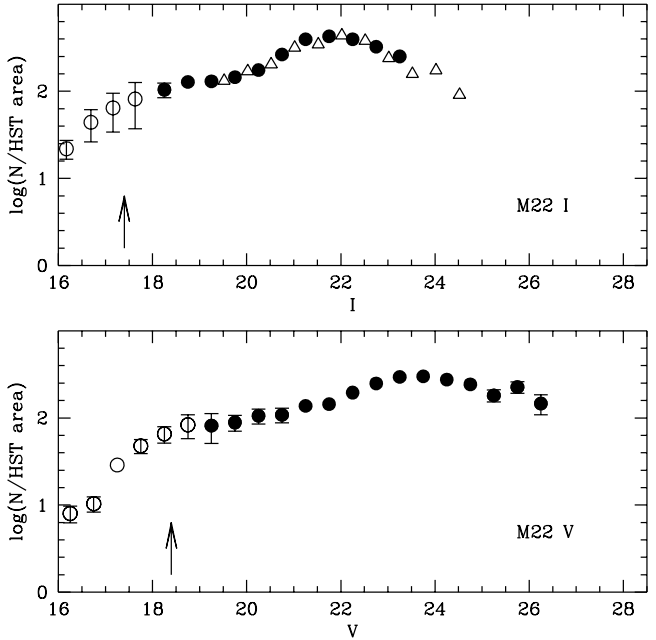}
\caption[]{The $V$ ({\it lower panel}) and $I$ ({\it upper panel}) 
ground-based ({\it open circles}) and HST ({\it filled circles}) LFs 
of M22 from the CMD of Fig.~\ref{cm22}. The ground-based LFs have been 
scaled to the HST area. The arrows indicate the TO position. The error 
bars are plotted only when they exceed the symbol size. The {\it open 
triangles} show the LF obtained by De Marchi and Paresce (1997) from 
the same HST data.}
\label{m22_gb_hst}
\end{figure}

\begin{figure}
\psfig{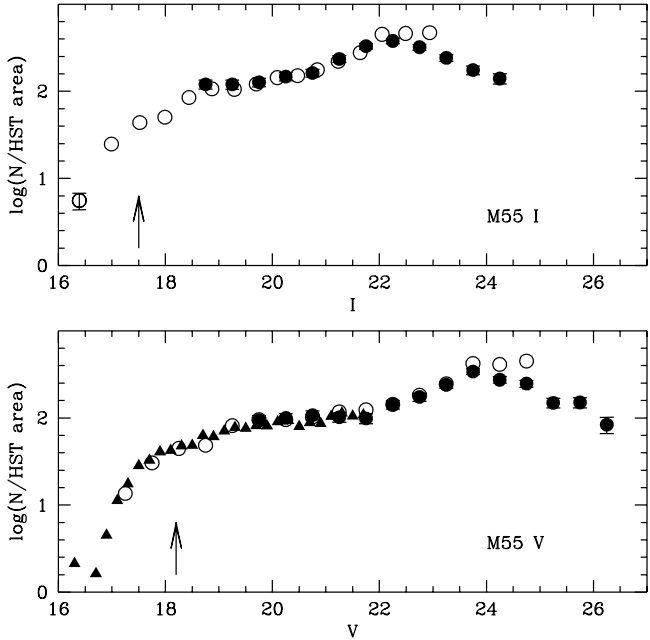}
\caption[]{The $V$ ({\it lower panel}) and $I$ 
({\it upper panel}) ground-based and HST ({\it filled circles}) LFs of M55
from the CMD of Fig.~\ref{cm55}. The ground-based LFs are from Zaggia
\etal\ (1997, {\it filled triangles}) and Mandushev \etal\ (1996,
{\it open circles}). The LF from Zaggia \etal\ have been normalized
to the HST area, while the Mandushev \etal\ LF has been scaled to the
total counts in the common magnitude range. The arrows indicate the TO 
position. The error bars are plotted only when exceeding the symbol size.}
\label{m55_gb_hst}
\end{figure}

Both the $V$ and $I$ LFs for the three clusters are shown in
Figs.~\ref{m10_gb_hst}, \ref{m22_gb_hst}, and \ref{m55_gb_hst} and
listed in Tables~\ref{m10lf}, \ref{m22lf}, and \ref{m55lf}.  Col.~1
gives the $V$ magnitudes, Col.~2 gives the field-star corrected LF,
and Col.~3 lists the corrected (for both field star contamination and
incompleteness) counts. The same figures for the $I$ LFs are in
Cols.~4--6. The LFs listed in Tables~\ref{m10lf}--\ref{m55lf} are from
the HST photometry up to the brightest magnitude bin of the HST data;
for brighter magnitudes Tables~\ref{m10lf}--\ref{m55lf} list the
ground-based LFs scaled to the area of one WF CCD.

The ground-based LFs in Figs.~\ref{m10_gb_hst} and \ref{m22_gb_hst} 
({\it open circles}) have been obtained from the CMDs of Figs.~\ref{cm10} and 
\ref{cm22}, respectively. As the ground-based and HST LFs ({\it filled circles})
have been obtained at the same distance from the cluster center, they
have been normalized by scaling the ground-based counts to the area of
the HST field. In Fig.~\ref{m55_gb_hst} we compare the HST LF of M55
with the LFs from Zaggia \etal\ (1997) and from Mandushev \etal\
(1996). Note that the data of Zaggia \etal\ refer to the same radial
interval as the HST data, while the field of Mandushev \etal\ is located in
an outer region (at 6.8 arcmin from the cluster center). In this case,
the LF has been normalized to the others by using the total star counts
in the common magnitude interval.  Note the good agreement among the
three LFs in the common region. The LF of Mandushev is
slightly steeper than the HST LF, particularly at the lower end; this
might be due to a mass-segregation effect.
In Fig.~\ref{m22_gb_hst} the {\it open triangles} show the LF obtained by
De Marchi and Paresce (1997) from the same HST data. Despite the fact that
the photometry has been obtained in a completely independent way, and using 
rather different reduction procedures, it is comfortable to 
see that there is perfect agreement between the two LFs down to $I=23.25$,
where the completeness is only 56\% (\cf\ Table~\ref{m22lf}). We have
already commented how, below $I\sim23.5$,
the incompleteness and, most importantly, the difficulties in the field-star 
contamination estimates make the LF rather uncertain, and we prefer to
avoid presenting data that are too uncertain.

As was discussed in Section~\ref{intro}, all three clusters have
comparable metallicities (within $\sim 0.3$ dex). This fact allows to
compare directly their LFs, without having to pass through the
uncertain MLRs. Similarities or differences among the LFs of the three
clusters reflect directly similarities or differences in their MFs.
In addition, the HST fields where the LF was measured for M10, M22,
and M55 are all located very close to the half-mass radius
($r\subr{obs}/r\subr{h}$=1.1 for M55, $r\subr{obs}/r\subr{h}$=1.4 for
M10, and $r\subr{obs}/r\subr{h}$=1.8 for M22). This is a particularly
fortunate case, as Vesperini and Heggie (1997), among others, have
shown that the LF observed close to the half-mass radius in a King
model (i.e., not collapsed) cluster, is close to the global (present
day) LF. In other words, as a first approximation, our LFs do not need
any mass-segregation correction.


\begin{table}
\caption[]{M10 Luminosity Function}
\label{m10lf}
\begin{tabular}{llllll}
\hline
$V$ & $N$ & $N_{corr}$ & $I$ & $N$ & $N_{corr}$ \\
\hline
    18.25   &   9   &  11  &  16.75  &   4 &    4  \\ 
    18.75   &  16   &  20  &  17.25  &   6 &    9  \\
    19.25   &  21   &  27  &  17.75  &  11 &   20  \\
    19.75   &  25   &  32  &  18.25  &  16 &   30  \\
    20.25   &  53   &  46  &  18.75  &  41 &   34  \\
    20.75   &  72   &  68  &  19.25  &  75 &   64  \\
    21.25   &  87   &  94  &  19.75  &  98 &   98  \\
    21.75   &  89   &  78  &  20.25  & 115 &  115  \\
    22.25   &  79   &  68  &  20.75  & 101 &   99  \\
    22.75   & 108   & 110  &  21.25  & 149 &  157  \\
    23.25   & 116   & 124  &  21.75  & 179 &  202  \\
    23.75   & 163   & 187  &  22.25  & 222 &  255  \\
    24.25   & 188   & 220  &  22.75  & 250 &  323  \\
    24.75   & 203   & 260  &  23.25  & 180 &  243  \\
    25.25   & 175   & 230  &  23.75  & 141 &  195  \\
    25.75   & 126   & 167  &  24.25  & 104 &  150  \\  
    26.25   & 102   & 161  &  24.75  &  46 &   86  \\  
    26.75   &  55   &  75  &         &     &       \\
    27.25   &  45   &  80  &         &     &       \\
    27.75   &  19   &  34  &         &     &       \\
\hline
\end{tabular}
\end{table}

\begin{table}
\caption[]{M22 Luminosity Function}
\label{m22lf}
\begin{tabular}{llllll}
\hline
$V$ & $N$ & $N_{corr}$ & $I$ & $N$ & $N_{corr}$ \\
\hline
    15.25   &     6	&   6 &	13.94   &    5	&   5	\\
    15.75   &     5	&   5 &	14.46   &    5	&   5	\\
    16.25   &     8	&   8 &	14.99   &    7	&   7 	\\
    16.75   &    10	&  10 &	15.54   &    8	&   8	\\
    17.25   &    29	&  29 &	16.17   &   22  &  22	\\
    17.75   &    40	&  48 &	16.69   &   44	&  36	\\
    18.25   &    51	&  65 &	17.16   &   65  &  51	\\
    18.75   &    59	&  83 &	17.63   &   81	&  58	\\
    19.25   &    76    	&  82 &	18.25   &  105	& 104	\\
    19.75   &    89   	&  89 &	18.75   &  127	& 128	\\
    20.25   &   120  	& 106 & 19.25   &  119	& 130	\\
    20.75   &   111  	& 108 &	19.75   &  123	& 145	\\
    21.25   &   141 	& 137 &	20.25   &  144	& 175	\\
    21.75   &   149  	& 144 &	20.75   &  197	& 264	\\
    22.25   &   185 	& 195 &	21.25   &  292	& 395	\\
    22.75   &   215  	& 249 &	21.75   &  272	& 427	\\
    23.25   &   244  	& 295 &	22.25   &  237	& 397	\\
    23.75   &   245  	& 301 &	22.75   &  189	& 325	\\
    24.25   &   197  	& 276 &	23.25   &  140	& 251	\\
    24.75   &   164  	& 243 &	        &       &       \\
    25.25   &   125 	& 181 &	        &       &       \\
    25.75   &   121  	& 226 &	        &       &       \\
    26.25   &    77 	& 146 &	        &       &       \\
\hline
\end{tabular}
\end{table}

\begin{table}
\caption[]{M55 Luminosity Function}
\label{m55lf}
\begin{tabular}{llllll}
\hline
V & N & $N_{corr}$ & I & N & $N_{corr}$ \\
\hline

    19.75   &   89 &  94  & 18.75  &  96  & 120 \\
    20.25   &   93 & 100  & 19.25  & 109  & 120 \\
    20.75   &   89 & 107  & 19.75  & 109  & 127 \\
    21.25   &   95 & 101  & 20.25  & 135  & 148 \\
    21.75   &   96 &  98  & 20.75  & 142  & 162 \\
    22.25   &  132 & 142  & 21.25  & 189  & 235 \\
    22.75   &  151 & 173  & 21.75  & 255  & 330 \\
    23.25   &  201 & 240  & 22.25  & 285  & 378 \\
    23.75   &  256 & 341  & 22.75  & 224  & 321 \\
    24.25   &  229 & 274  & 23.25  & 158  & 242 \\
    24.75   &  166 & 247  & 23.75  & 108  & 175 \\
    25.25   &  115 & 149  & 24.25  &  73  & 140 \\ 
    25.75   &   99 & 150  &        &      &     \\ 
    26.25   &   57 & 83  &        &      &     \\ 
\hline
\end{tabular}
\end{table}

\begin{figure}
\psfig{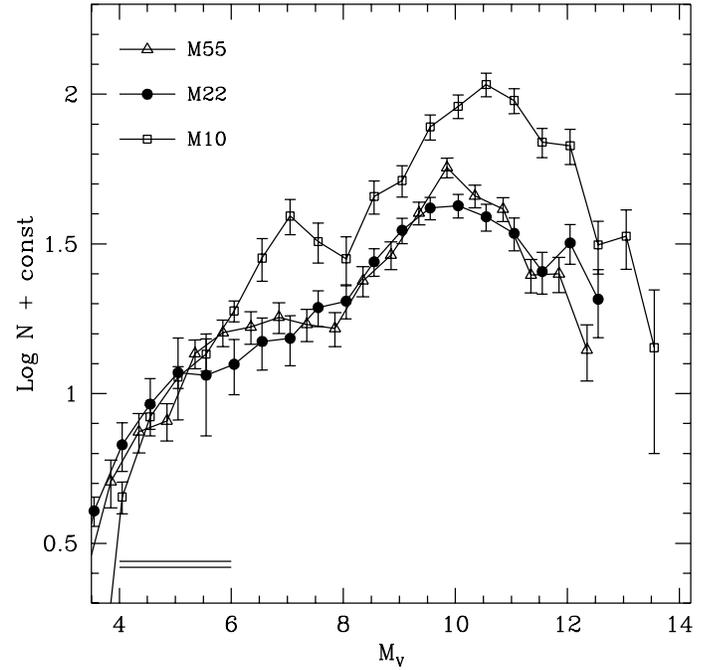}
\caption[]{The $V$ LFs of M10, M22, and M55. The three LFs have been 
extended to the TO using the ground-based data.  The horizontal bars 
indicate the normalization interval. The LFs of M10 is steeper than the
other two. The difference in slope is significantly higher than what expected
from the stellar structure models.}
\label{comp3v}
\end{figure}
\begin{figure}
\psfig{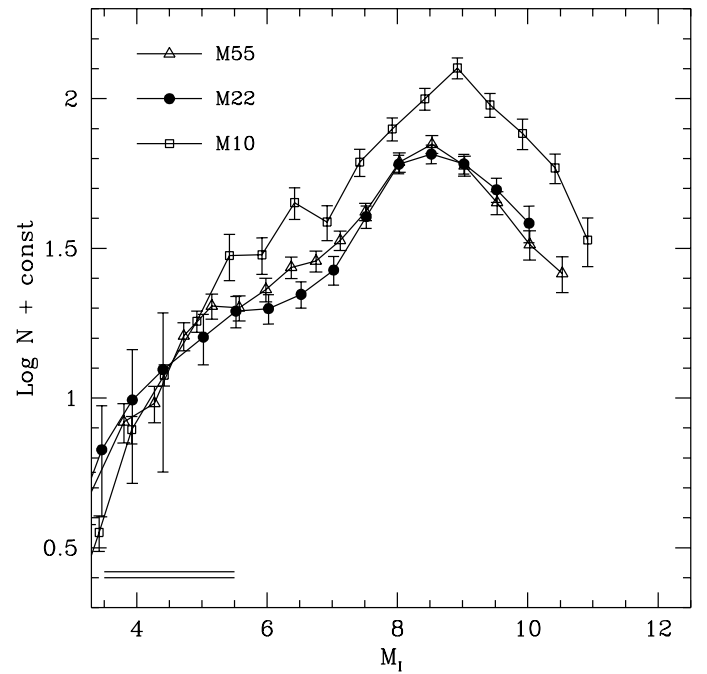}
\caption[]{As in Fig.~\ref{comp3v}, but for the $I$ LFs. Again the LF of M10
is significantly steeper.}
\label{comp3i}
\end{figure}

The comparison is shown in Fig.~\ref{comp3v} ($V$ LFs) and
Fig.~\ref{comp3i} ($I$ LFs). The adopted apparent distance moduli and
reddenings are given in Table~\ref{distance} (Baraffe \etal\ rows). In
the absence of a means of normalizing the three LFs to a global
cluster parameter, arbitrary constants determine the vertical
positioning of the individual LFs.  We have chosen these constants
exactly as described in PCK.  Briefly, vertical shift of the M10, M22,
and M55 LFs were made to bring them into alignment,
according to a least-square algorithm, in the magnitude intervals
$4.0<M\subr{V}<6.0$ and $3.5<M\subr{I}<5.5$. The overall trend of the
LFs in Fig.~\ref{comp3v} and Fig.~\ref{comp3i} is similar, with a steep
rise up to $M_V\sim10$ ($M_I\sim8.5$), followed by a drop to the
limiting magnitude of the present investigation. Indeed, both the $V$
and $I$ LFs for M10 reach their maximum at magnitudes $\sim0.5$
fainter than in M55 and M22. This is qualitatively consistent with the
fact that M10 is slightly more metal rich than the other two clusters
(D'Antona and Mazzitelli, 1995).

Despite the fact that the overall shape of the LF of M10 is similar to
the others, a close inspection of Figs.~\ref{comp3v} and \ref{comp3i}
reveals that it is significantly steeper than the other two LFs. This
difference can hardly be due to any internal dynamical evolution, if
we consider that the three clusters have similar internal
structures. Even if we did not fully trust the dynamical models (both
King-Michie models and N-body simulations, Vesperini and Heggie 1997),
which predict that our local LFs for M10, M22, and M55 closely
resemble the global ones, it is noteworthy that the M10 field is at an
intermediate position in terms of half-light radius $r_h$, between the
M55 and the M22 fields. Therefore, the differences in LF slopes cannot
be due to mass segregation.

\section{The mass functions} \label{mf}

\begin{figure}
\psfig{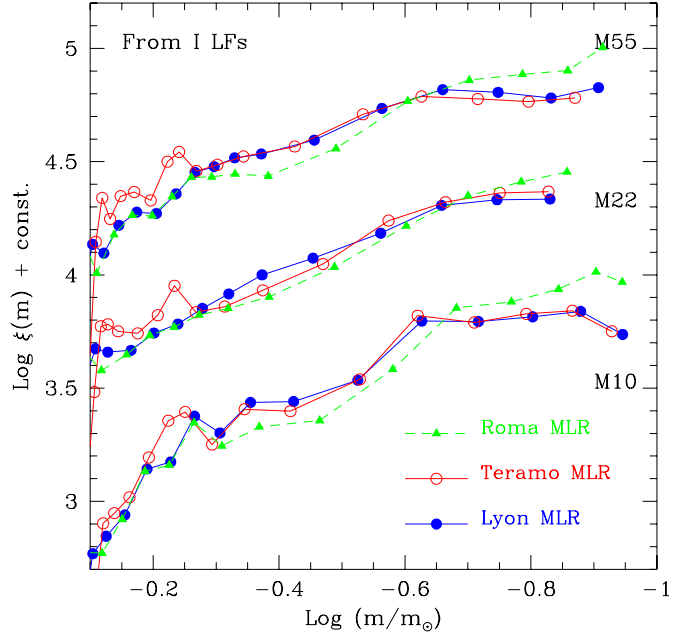}
\caption[]{The MFs of M10, M22, and M55 from the $I$ LFs and from
different theoretical MLRs are compared. For reasons of clarity the MFs
are arbitrarily shifted on the vertical axis.  }
\label{mfi}
\end{figure}

\begin{figure}
\psfig{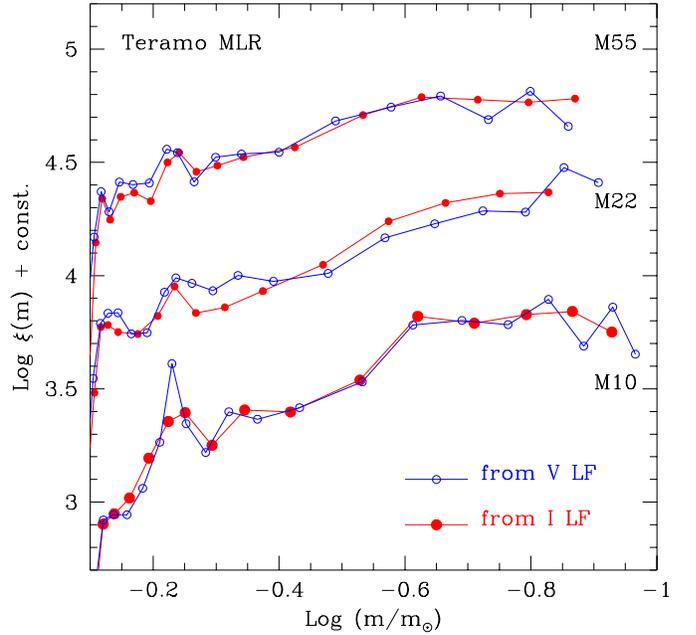}
\caption[]{Comparison of the MFs of M10, M22, and M55 from the $V$ and 
$I$ LFs and from the Teramo MLRs. For reasons of clarity the
MFs are arbitrarily shifted on the vertical axis.  }
\label{mfvi}
\end{figure}

The LFs can be transformed into MFs using a MLR.  As emphasized in
Section~\ref{cmd}, such transformations are still uncertain for low-mass,
low-metallicity stars, and we must rely on the models almost entirely.
It is somehow reassuring that not only at least two of the existing
models are able to reproduce the observed diagrams, 
but also the distance moduli and
reddenings that result from the fit are in agreement, within the
errors, with the values in the literature. This does not mean that the
models by the Lyon and the Teramo groups used in Section~\ref{cmd} are
the {\it correct} ones.  They are simply the best ones presently
available, and we will use both of them to gather some information on
the general shape of the MFs of M10, M15, and M22. The cautionary
remarks on the absence of empirical MLR data should still be heeded.  For
the sake of comparison, and in order to give an idea of the possible range of
uncertainty, we will also use the MLR of the Roma group 
(D'Antona and Mazzitelli 1995).

In Fig.~\ref{mfi} we compare the MFs derived from the $I$ LFs of M10,
M22, and M55 from the TO down to $\sim 0.11m_\odot$.  An arbitrary
vertical shift is applied for reasons of clarity.  The adopted distance
moduli, reddenings, and metallicities for the MFs obtained using the
Lyon and Teramo models are in Table~\ref{distance}.  For the Roma
model we adopted the values in Djorgovski (1993) and used the
isochrone corresponding to 10 Gyr.

The MFs obtained using the Lyon and Teramo models track one other closely 
for $m<0.6 m_\odot$. The small differences for higher masses might be
due, at least in part, to the difference in the adopted ages.  The MFs
from the Roma models are systematically steeper, as already noted in
PCK.  Fig.~\ref{mfvi} compares the MFs obtained from the $V$
and $I$ LFs, using the Teramo models.  In all cases, the two MFs are
very similar, despite the fact that the two LFs have been independently
obtained. This result is also reassuring on the theoretical side,
showing the internal consistency of the models.

In all cases, there is a hint of flattening at the low-mass
end, but no sign of a drop-off.

As the transformation from the LF to the MF is the weakest part of the
present analysis, we prefer not to comment further on the detailed
structure of the MF.  
We note only that the slopes of the MFs below $0.5m_\odot$, using any
existing MLRs, are shallower than the $x=1$ slope ($x=1.35$ for the
Salpeter MF in this notation) for which the integration of the total
mass down to $m=0$ would diverge.

\section{Comparison with other clusters}~\label{comp}

\begin{figure}
\psfig{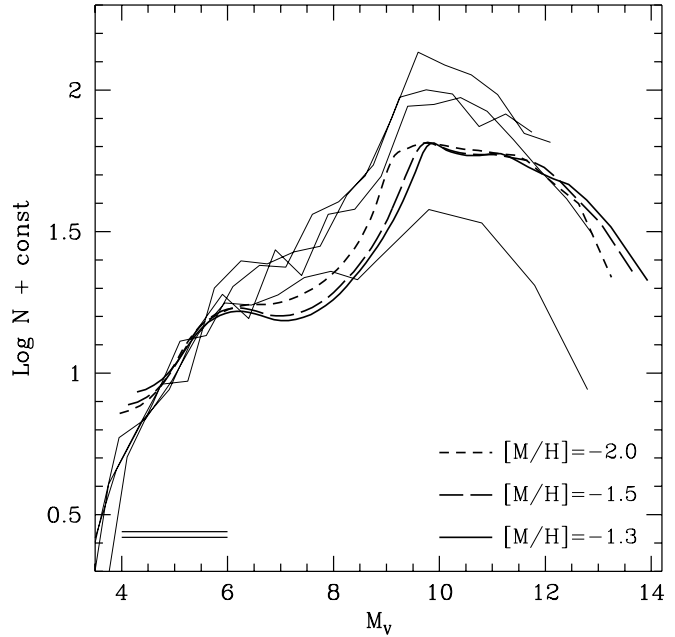}
\caption[]{The theoretical LFs from the Baraffe \etal\ (1997) models 
for a power law MF with slope x=-0.3 and metallicities as shown in the labels
are compared with the V LFs on M15, M30, M92, and NGC 6397. 
In the metallicity interval covered by the clusters in our sample,
the theoretical LFs look quite similar.}
\label{met}
\end{figure}

It is interesting to compare the LFs in Fig.~\ref{comp3v} and 
\ref{comp3i} with the LFs of other GCs with similar metallicity. 
The only other homogeneous set
of $V$ and $I$ LFs extending from the TO to $m<0.15m_\odot$ has been
collected by PCK for M15, M30, M92, and NGC~6397. In
addition, Ferraro \etal\ (1997) have published an $I$ LF for NGC~6752.
There are two other clusters, which have a metallicity comparable to M22
and M10, and for which deep HST LFs have been published:
$\omega$ Cen (Elson, Gilmore, and Santiago 1995) and M3 (Marconi
\etal\ 1998). 
In both cases, saturation of the brightest stars does not allow 
to extend the LFs to the TO and we will omit them in the present 
comparison.

\begin{figure}
\psfig{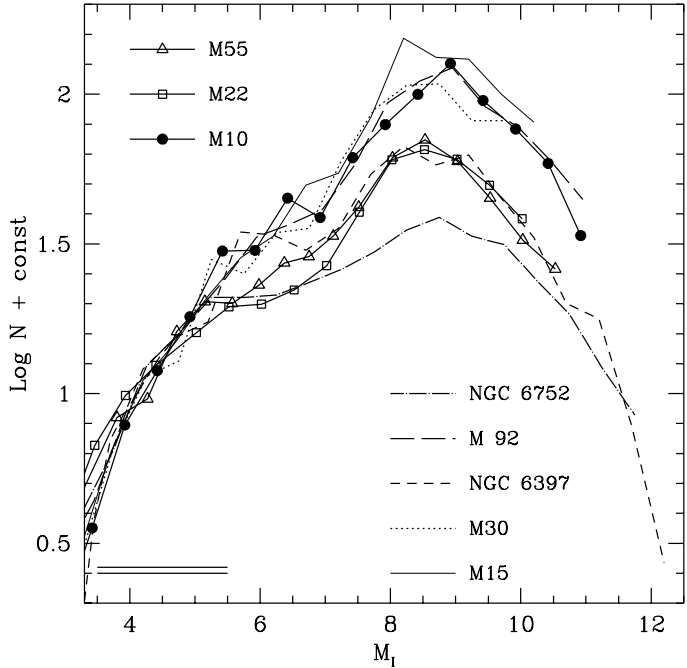}
\caption[]{As in Fig.~\ref{allv}, but for the $I$ LFs.
The $I$ LF of NGC~6397 is from King \etal\ (1998). 
The LF of NGC~6752 is from Ferraro \etal\ (1997).
As in Fig.~\ref{allv}, note the spread in slope.}
\label{allid}
\end{figure}

\begin{figure}
\psfig{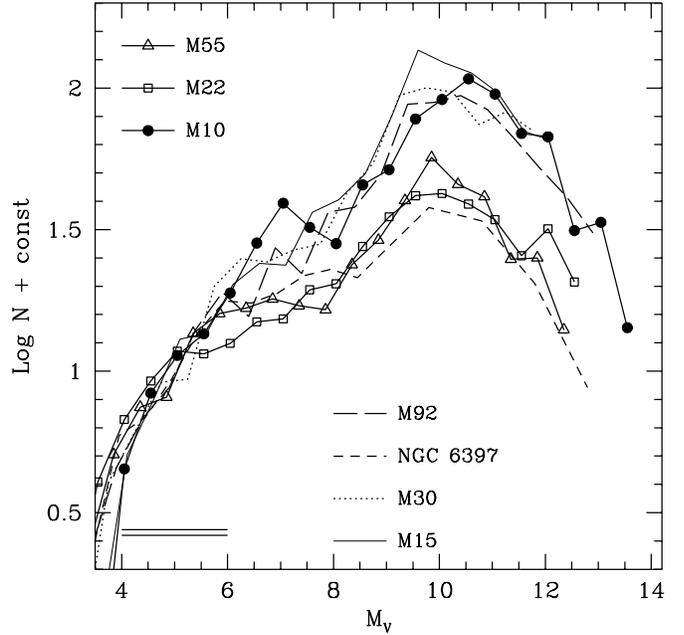}
\caption[]{
The $V$ and $I$ LFs of M10, M22, and M55 are compared with the LFs of
all the clusters with [Fe/H]$<-1.6$ available in the literature.  
Only the LFs which extend to the TO are used. 
The LFs for M15, M30, M92, and NGC 6397 are from PCK
The horizontal bars show the adopted normalization interval. 
Note the spread in slope of these LFs,  significantly larger than expected
from the error bars (\cf\ Fig.~\ref{comp3v}) and the metallicity differences
(\cf\ Fig.~\ref{met}). }
\label{allv}
\end{figure}

As discussed in King \etal\ (1995) and PCK, at the intermediate radius
at which M15, M30, M92, and NGC~6397 were observed, their LFs are
fortuitously close to the global ones, with differences that nowhere
exceed a few tenths in the logarithm (PCK), with the local LFs being
always steeper than the global ones.  So, a comparison of these LFs
with the LFs of M10, M22, and M55 should be only marginally affected
by mass segregation effects.  Unfortunately, no detailed dynamical
model for NGC~6752 is available at the moment.  We ran a multi-mass
King-Michie model on this cluster (\cf\ Section~\ref{disc}). 
We find that the locally observed
LFs of NGC~6752 is significantly flatter than the global one. 

\begin{figure}
\psfig{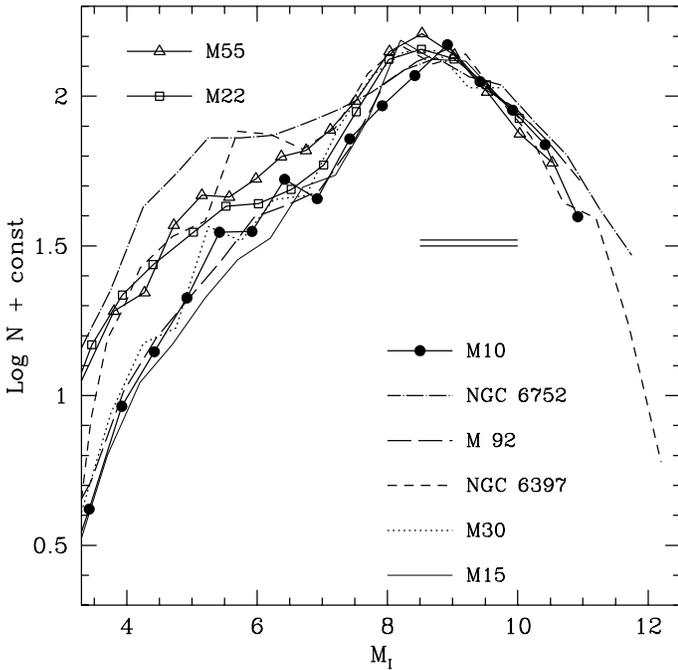}
\caption[]{As in Fig.~\ref{allid}, but with a different normalization 
interval.}
\label{alliu}
\end{figure}

The metallicity of the three clusters presented in this paper is
slightly higher than that of the clusters in PCK.  We
need to investigate the effect of this metallicity spread on the LFs.
In Fig.~\ref{met} the theoretical LFs for a power law MF with a slope
x=0.3 and three different metallicities are plotted together with the
$V$ LFs of the four clusters in PCK used as reference.
The three theoretical LFs are quite similar in the metallicity
interval which span the entire metallicity range of the clusters
discussed in the following.

Figs.~\ref{allv} and \ref{allid} compare the $V$ and $I$ LFs of M10,
M22, and M55 with the LFs of all the clusters with [Fe/H]$<-1.6$
available in the literature. Only the LFs which extend to the TO are
used.  The overall shape of the LFs is similar for the eight clusters,
with a steep rise up to $M_V\sim10$ ($M_I\sim8.5-9.0$), followed by a
drop to the limiting magnitude.  As shown in Fig.~\ref{met} this shape
is mainly due to the mass-luminosity relation (MLR) and not to the
morphology of the MF (\cf\ also Section~\ref{mf}).  Despite the fact
that the general trend of these LFs is similar, their slope is
different.  These differences are much larger than expected from the
error bars (Figs.~\ref{comp3v} and \ref{comp3i}) and from
the cluster metallicity differences (Fig.~\ref{met}), i.e. imply
different (local) MFs. In view of the small expected correction for 
the mass segregation above discussed, this also imply 
different global present day MFs.

Had we chosen to align the LFs at the faint end instead of the bright end, 
the result would have been the same, as shown in Fig.~\ref{alliu}.
This figure explicitly shows that the differences among the GC LFs become
apparent only when they are extended to the TO.

\section{Discussion} \label{disc}

The differences among the LFs noted in the previous Section suggest
differences among the observed MFs. They mainly imply that the ratio
of the low mass to high mass stars differs from cluster to cluster. 
It is interesting to further comment on the possible origin
of these differences. There are two possibilities:

\begin{itemize}

\item The differences are primordial, i.e. the Galactic GCs are born 
with different IMFs;

\item The differences are a consequence of the dynamical evolution, both
internal (energy equipartition, evaporation), or externally induced
by the gravitational potential of our Ga\-la\-xy.

\end{itemize}

Of course, it is not possible to test the first hypothesis directly,
and a combination of the above two possibilities cannot be excluded.
PCK, relying on the presently available models for M15, M30, M92, and
NGC~6397, and on the Galactic orbits of some of these clusters,
propose that the different shape of the LF of NGC~6397 with respect to
the other three arises from the interaction of this cluster with the
Galaxy. It is a consequence of the frequent tidal shocks experienced
by NGC~6397 along its orbit.

At the time we are writing this paper, 
the number of clusters at our disposal is more than doubled,
though it is still too small for any detailed analysis.  As discussed
in the previous Section, among the GCs with deep HST LFs, there are
8 clusters with a LF extending from the TO to somewhat above 0.1
$m_\odot$.  As a sort of exercise, we can try to look for possible
dependences of the overall shape of the MFs on the observable
parameters which characterize the GGC population.

This analysis can be done if we can:

\begin{itemize}

\item Correct for mass segregation effects (though these have
been anticipated to be small, with the only exception of \\
NGC~6752);

\item Parametrize in some way the MF shape.

\end{itemize}

In view of the still relatively small sample of objects, we have
chosen a very simple approach.  We have limited our analysis to the 3
clusters presented in this paper, the four objects of PCK, and
NGC~6752 (Ferraro \etal\ 1997). The other three objects with deep HST
LFs have not been considered as their LFs are not complete, lacking the
brightest part (from $\sim 0.5$ m$_\odot$ to the TO).

First of all, the LFs have been trasformed into MFs using Baraffe et
al.'s (1997) MLRs for the appropriate metallicity and using the
distance modulus which best fits the CMDs (\cf\ previous Sections).

Second, we run a King-Michie model in order to have a first
approximation correction for the mass segregation effects.  We used
the code kindly provided by Jay Anderson and described in Anderson
(1998), which is based on the Gunn and Griffin (1979) formulation of
the multimass King--Michie model. These models have a
lowered-Maxwellian distribution function, which approximate the
steady-state solution of the Fokker--Planck equation (King 1965). In
the case of the post--core--collapse (Djorgovski and King 1986) 
clusters these models do not
incorporate important physical effects, most importantly, the
deviation from a Maxwellian distribution function in the collapsed
core (Cohn 1980). However, they are the simplest models that can
predict the radial variation of the MF due to energy equipartition,
and rather realistically, as shown by King \etal\ (1995) and Sosin
(1997, see also King 1996). On the other side, as shown by Murphy et
al. (1997) for M15, mass segregation prediction of King-Michie models
are not very different from what found with more sophisticated
Fokker--Planck models.

For NGC~6397 we used the same model parameters obtained by King \etal\
(1995), for M15 the parameters by Sosin and King (1997), and for
M30 the parameters by Sosin (1997). For M92, we used the model
parameters by Anderson (1998). The details of the models for the other
clusters will be described elsewhere.  We essentially followed the
same procedure described in Sosin (1997). Briefly, we calculated the
model by first choosing the core and tidal radii given by Trager,
King and Djorgovski (1995).  We then defined 17 mass groups, whose
number of stars and averaged masses were constrained to agree with the
observed MF at the distance from the cluster center of the observed
field.  We then added a group of 0.55 $m_\odot$ white dwarfs chosen
(somewhat arbitrarely) to contain 20\% of the cluster mass. Finally,
we added 1.4 $m_\odot$ dark remnants and adjusted their mass fraction
(usually around 1.5\%) to make the radial profile of the stars in the
brightest bin agree with the surface density profile of Trager et
al.  (1995).

One of the outputs of the models are the global MFs for each
cluster. In our model, the NGC~6752 field is located in a position 
strongly affected by mass segregation. In view of the large correction 
we should apply to its local MF in order to have the global one, the further 
uncertainties in the model due to the post--core--collapse status of this cluster,
and the strong deviation from a power law of its MF (likely a further
effect of the mass segregation), we will not include NGC~6752
in the following analysis. 

We fitted the global MFs of the remaining seven clusters with a power law
$\xi(m)=\xi_0m^{-(1+x)}$, and used the index $x$ as a parameter
indicating the ratio of low mass to high mass stars. The power law has
proven to fit reasonably well all the seven global MFs.
In view of the uncertainties associated to the MLR and to the model
itself, any more sophisticated analysis is not justified.

\begin{figure*}
\psfig{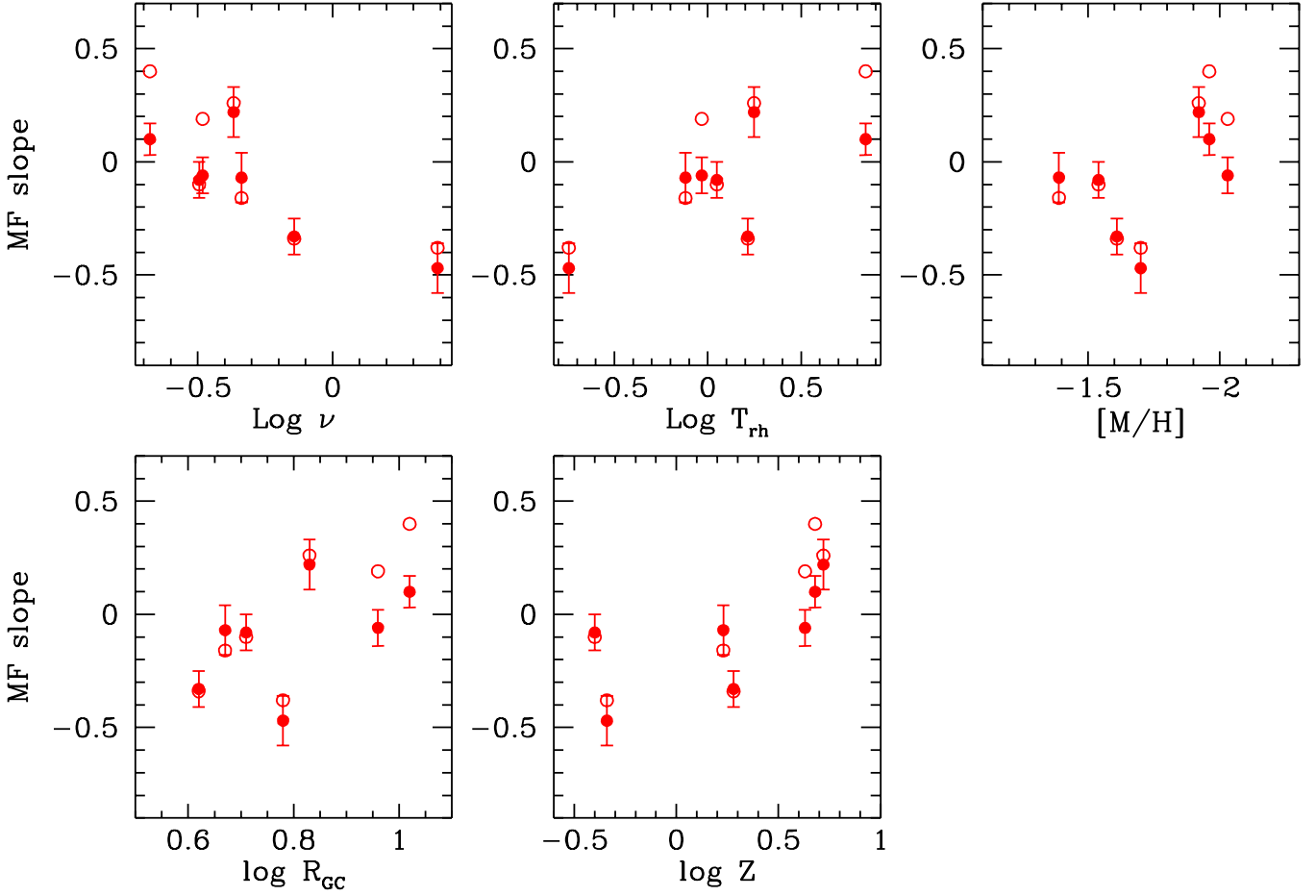}
\caption[]{Univariate correlations of the slopes of the {\it global}
({\it full dots}) and {\it local} ({\it open circles}) MFs with the
distance from the Galactic center $R_{\rm GC}$, from the Galactic plane
$Z$,  the half mass relaxation time ($T_{\rm rh}$), the destruction rate $\nu$,
and the metallicity [M/H], calculated as in Section~\ref{cmd}, assuming 
[O/Fe]=0.3.  The MFs have been fitted with a power
law in the mass interval $m<0.7 m_\odot$. The error bars represent the
standard deviation of the slope of the straight line fitted to the data
in the Log-Log plane. Note the large dispersion of the data. Clusters
with smaller $T_{\rm rh}$ and $\nu$ have flatter MFs.}
\label{mfslopes}
\end{figure*}

The slopes of the global present day MFs for $m<0.7m_\odot$ for the
clusters in our sample are plotted in Fig.~\ref{mfslopes} against the
half mass relaxation time ($T_{\rm rh}$), the position within the
Galaxy (the galactocentric distance $R_{\rm GC}$ and the distance from
the disk $Z$) and the destruction rates $\nu$ (in units of inverse Hubble time)
as calculated by Gnedin and Ostriker (1997). 
The error bars show the formal 
error of the fit, and do not include the error in the MLR and the error
associated to the mass segregation correction. The {\it full circles} show the
slope of the {\it global} MF, while the {\it open circles} refer to the slopes of the
original ({\it local}) MF. As already anticipated, the corrections for
mass segregation are in general small. The discussion which follows applies to
both the global and local MFs.

The slopes have a large dispersion, showing
that the present day global MFs significantly differ from cluster to cluster.

Clusters with larger $\nu$ (and smaller $T_{\rm rh}$) tend to have
flatter MFs, suggesting that the observed differences in the MF slopes
might be related to the cluster dynamical evolution.  There is also an
indication of a trend with the distance from the Galactic center which
resembles a similar dependency suggested by Djorgovski, Piotto and
Capaccioli (1993), and which has been interpreted in terms of
evidences of a dynamical evolution (Capaccioli, Piotto and Stiavelli
1993).  Also a dependency on metallicity cannot be excluded, as shown
in Fig.~\ref{mfslopes}.  However, the uncertainty associated to the
various transformations from the local LFs to the global MFs, and the
small number of points do not allow to assess the significance of any
of these trends.

Another interesting feature from the slopes in Fig.~\ref{mfslopes} is
their low value. For $m<0.7 m_\odot$ the MFs never exceed a slope
$x=0.3$, ranging in the interval $-0.5<x<0.3$ (in the scale in which
the Salpeter MF has a slope $x=1.35$). Also the field star MF is
significantly flatter than the Salpeter MF in the low mass regime
(Tinney 1998).  Mera, Chabrier, and Baraffe (1996) find a slope $x=1$
for the disk MF for $m<0.6 m_\odot$, and Scalo (1998) in his review
concludes that an average slope $x=0.2$ is appropriate for
$0.1<m/m_\odot<1.0$ for both Galactic clusters and field stars. A
slope $x=0.3$ is also proposed by Kroupa, Tout, and Gilmore (1993) and
Kroupa 1995 for the field stars.  More recently, Gould, Bahcall, and 
Flynn (1996) find a
smaller $x=-0.1$ for $m<0.6m_\odot$, though this value is uncertain
because it is based on a small number of stars. Still, the
Galactic cluster and field star MF slopes seem to represent an {\it
upper limit} to the GC present day MF slopes.

This result implies that either the IMF of some GCs was flatter than
the field MF, or the present day MFs in GCs do not represent the
IMF(s), at least for some of the clusters in this sample. This could
imply an evolution of the GC MF with time, with a tendency to
flatten. This might be another evidence of the presence of dynamical
evolution effects as predicted by the theoretical models (Chernoff and
Weinberg 1990, Vesperini and Heggie 1997) and invoked by Capaccioli et
al. (1993) and PCK in order to explain the observed GC MFs.

Another consequence of the flatness of the MF is that the contribution
to the total cluster mass by very-low-mass stars and brown dwarfs is
likely to be negligible. Of course, the fraction of mass in form of
brown dwarfs depends on (1) the MF slope in the corresponding mass
interval and (2) the lower limit we assume for the brown dwarf masses 
($m_{\rm low}$).  As a working hypothesis, we might assume that we can extrapolate
the MF power law which represents as a first approximation the GC present day MF
for $0.1<m/m_\odot<0.8$ to the brown dwarf regime. We also arbitrarely
adopt $m_{\rm low}=0.01 m_\odot$, which corresponds to the minimum
Jeans mass at zero metallicity (Silk 1977).  Even for the steepest
slope ($x=0.3$) in Fig.~\ref{mfslopes}, stars with $m_{\rm low}<m/m_\odot<0.09$
($m\sim0.09 m_\odot$ correspond to the minimum mass for the core hydrogen
burning ignition) contribute to less than 20\% of the total mass of the
stars with $m<0.8 m_\odot$, though they are more than 55\% in number.
If the brown dwarf MF is not radically different (steeper) than the main
sequence MF, they cannot contribute in any significant way to the cluster
dynamics.

\begin{acknowledgements}
We are deeply grateful to Peter Stetson for providing us the mask
files for vignetting and pixel area correction for WFPC2, and the PSF
files, and for providing the most up-to-date versions of his programs.
We thank also Jay Anderson for making available his multimass
King-Michie code.  We are particularly indebted to Ivan King who
carefully read the manuscript and persuasively pushed us to make the
mass segregation corrections and the comparisons of the last Section.
This work has been supported by the Agenzia Spaziale Italiana and the
Ministero della Ricerca Scientifica e Tecnologica.

\end{acknowledgements}

\end{document}